# Pressure effects on the $T_C$ of superconducting $MgC_xNi_3$

H. D. Yang[1,†], S. Mollah[1,*], W. L. Huang[1], P. L. Ho[1], H. L. Huang[1], C. -J. Liu[2],
J. -Y. Lin[3], Y. -L. Zhang[4], R. -C. Yu[4], and C. -Q. Jin[4]

[1]*Department of Physics, National Sun Yat Sen University, Kaohsiung 804, Taiwan, R.O.C.*
[2]*Department of Physics, National Changhua University of Education, Changhua 500, Taiwan, R.O.C.*
[3]*Institute of Physics, National Chiao Tung University, Hsinchu 300, Taiwan, R.O.C.*
[4]*Institute of Physics, Center for Condensed Matter Physics and Beijing High Pressure Research Center,
Chinese Academy of Sciences, P.O. Box 603, Beijing 100080, P.R.C.*

The effect of hydrostatic pressure (P) up to 17 kbar on the superconducting transition temperature ($T_C$) of the newly discovered intermetallic non-oxide perovskite superconductor $MgC_xNi_3$ has been reported. The $T_C$ is found to increase with increasing P at a rate of $dT_C/dP \sim 0.0134$ to $0.0155$ K/kbar depending on the value of carbon content x. The absolute value of $dT_C/dP$ for $MgC_xNi_3$ is about the same as that of intermetallic $RNi_2B_2C$ (R = rare earths) and metallic superconductors but about one order of magnitude smaller than that of the most recently and intensively studied superconductor $MgB_2$. However, the $d\ln T_C/dP \sim 0.00181$ to $0.00224$ kbar$^{-1}$ and the rate of change of $T_C$ with unit cell volume (V), $d\ln T_C/d\ln V \sim -3.18$ to $-2.58$ of $MgC_xNi_3$ are having the comparable magnitude to that of $MgB_2$ with opposite sign. The increase of $T_C$ with P in $MgC_xNi_3$ can be explained in the frame work of density of states (DOS) effect.

Soon after the discovery of a record high $T_C$ (~ 39 K) intermetallic non-cuprate superconductor $MgB_2$,[1] a new intermetallic non-oxide superconductor $MgCNi_3$ was found[2] to undergo a superconducting transition at $T_C \sim 8$ K. Though the $T_C$ of $MgCNi_3$ is much lower than that of $MgB_2$, it still attracts a lot of attention due to at least having the following physical significance related to the present studies. (1) It has a perovskite structure as the 30 K oxide non-cuprate superconductor $Ba_{1-x}K_xBiO_3$.[2] (2) A high proportion of Ni in this compound indicates that the magnetic interactions may play a dominant role to understand its superconductivity. (3) Its normal state NMR properties are irregular[3] and analogous to that observed in the exotic superconductor $Sr_2RuO_4$. (4) A typical isotropic *s*-wave superconductivity[3] is displayed by the nuclear spin-lattice relaxation rate $(1/T_1)$ with a coherence peak below $T_C$. (5) The change from grain boundary to core pinning by intragranular nanoparticles near $T_C$ proposes that the arrangement of pinning sites in $MgCNi_3$ is unique.[4] (6) The Hall coefficient and thermoelectric power data[5-6] show that the carriers in this compound are electrons in contrast to $MgB_2$. (7) Energy band calculations[7-9] demonstrate that the density of states (DOS) of the Fermi level ($E_F$) is dominated by Ni *d* states and there is a von Hove singularity (*vHs*) of the DOS just below (< 50 meV) the $E_F$.[8] Moreover, the photoemission and x-ray absorption studies show that the sharp *vHs* peak theoretically predicted near $E_F$ is substantially suppressed which may be due to electron-electron and electron-phonon interactions.[10]

It is well known that the high pressure (P) plays an important role on the $T_C$ of the metallic and intermetallic superconductors.[11-19] In general, the P can change the electronic structure, phonon frequencies or electron-phonon coupling that affecting the $T_C$. Both positive and negative pressure derivatives, $dT_C/dP$, are observed in the metallic and intermetallic superconductors.[11-19] For example, simple *s,p,d*-metal superconductors[16] like Sn, In, Ta or Hg, and the intermetallic superconductor like the recently discovered $MgB_2$ (Refs. 11-13) have shown the decrease of $T_C$ with the increase of P. However, depending on the rare earth site of the quaternary borocarbides, $RNi_2B_2C$ (R = rare earths), both increase and decrease of $T_C$ are observed with the increase of pressure.[14-15] In addition, the pressure can basically shift the Fermi level ($E_F$) towards higher energies[14-15] and thereby provide a probe on the slope of the DOS near $E_F$. Moreover, it can also modify the magnetic pair breaking effect and tune the competitive phenomena between superconductivity and spin fluctuations. From our magnetic field dependent resistivity and specific-heat studies,[20-21] it has been suggested that the $MgCNi_3$ is basically a typical BCS-like superconductor. In this report, we further present the pressure effects on the $T_C$ of this exotic superconductor to testify the mentioned unique electronic and magnetic properties.

The details of $MgC_xNi_3$ samples preparation and characterization can be found in Ref. 2 and 22. With increasing the nominal carbon content, the $T_C$ was improved. Depending on the values of nominal carbon x, the samples with different $T_C$'s are hereafter referred as A(x = 1.0), B(x = 1.25) and C(x

= 1.5). Electrical resistivity (ρ) of $MgC_xNi_3$ was measured by the standard four-probe method. Thermoelectric power (S) measurements were performed with the steady state techniques. The hydrostatic pressure (P) dependent ac magnetic susceptibility ($\chi_{ac}$) data were taken by the piston cylinder self-clamped technique.[23] The hydrostatic pressure environment around the sample was generated inside a Teflon cell with 3M Fluroinert FC-77 as the pressure transmitting medium. The pressure was determined by using Sn manometer situated near the sample in the same Teflon cell. In each instance, the original value was reproduced within experimental error after the pressure released indicating complete reversibility of the pressure effect.

Figure 1 shows the temperature dependence of resistivity (ρ) and thermoelectric power (S) for sample A. The inset of Fig.1 displays the ρ of samples A, B and C near $T_C$. The variation of ρ with temperature shows the same trend as reported in the literature[2,5-6,24-25] with $T_C$ ~ 7-8 K, $\rho_{300K}/\rho_{10K}$ ~2.3 and 90%-10% transition width ~0.2 K. The different value of $T_C$ for three studied $MgC_xNi_3$ samples is mainly due to the carbon stoichiometry.[2,24] The temperature dependence of S is negative confirming the carriers to be electron type which is consistent with the published results.[5-6] The nonlinear temperature dependence of S seems to suggest that the enhancement of electron-phonon interaction plays an important role in the superconductivity of $MgC_xNi_3$ like in cheverl-phase compounds[26] $Cu_{1.8}Mo_6S_{8-y}Se_y$ and $Cu_{1.8}Mo_6S_{8-y}Te_y$.

Temperature variation of ac magnetic susceptibility ($\chi_{ac}$) of samples A, B and C under pressure (0-17 kbar) is shown in Fig. 2. At ambient pressure, $T_C$ (~ 6.5 K) of sample A is the same as that obtained from specific heat[21] but little lower than that from the resistivity data (Fig. 1). The $T_C$ (midpoint) for sample A increases from 6.56 to 6.79 K with the increase of pressure from ambient to 14.80 kbar as shown in Fig. 3 having the rate of $dT_C/dP$ ~ 0.015 K/kbar and $dlnT_C/dP$ [= $(1/T_C)(dT_C/dP)$] ~ 0.002 $kbar^{-1}$. The similar trend of pressure effect on $T_C$ for samples B and C is also shown in Figs. 2 and 3. The positive values of $dT_C/dP$ and $dlnT_C/dP$ for these three samples are listed in Table 1. It is noted that these values of $dlnT_C/dP$ (Table 1) for $MgC_xNi_3$ lie on the range of ~0.001-0.008 $kbar^{-1}$ of conventional superconductors.[27]

For a clear and detailed idea of the pressure effect on the $T_C$ of other metallic and intermetallic superconductors, some of them are also listed in Table 1 for comparison. The decrease and increase of $T_C$ are observed respectively in metallic superconductors Ta and V with analogous magnitude of $dT_C/dP$ and $dlnT_C/dP$ as $MgC_xNi_3$ (Table 1). It is explained by the decrease of electron-phonon coupling constant in Ta and by the suppression of spin fluctuations as well as the increase of electron-phonon coupling in V.[16-18] The magnitude of positive $dT_C/dP$ for electron-carrier $MgC_xNi_3$ is about the same as that of its three-dimensional analogue $LuNi_2B_2C$ superconductor, and the latter one has been interpreted by the increase of DOS with P.[14] However, the negative $dT_C/dP$ and $dlnT_C/dP$ for hole-carrier $MgB_2$ may be either by a decrease of DOS (Ref. 12) or by a lattice stiffening.[13]

The change of $T_C$ with the unit cell volume (V) can be given by[11,14]

$(V/T_C)(dT_C/dV)=dlnT_C/dlnV=-(B/T_C)(dT_C/dP)$, (1)

where $B$ is the bulk modulus of the superconductor. Using the calculated value of $B$ for $MgC_xNi_3$ as 1510 kbar (Ref. 25) and taking the obtained $dT_C/dP$ and $T_C$ from Table 1, the $dlnT_C/dlnV$ values are found from Eq. (1) respectively for samples A, B and C as –3.18, –2.58, and –2.76. These values are of the same order of magnitude in $MgB_2$ superconductor (+ 4.16) with opposite sign.[11]

Since the DOS is sufficiently large in $MgC_xNi_3$ to produce strong electron-phonon coupling[9] and is supported by its S data, the $T_C$ can be expressed by the McMillan formula[28] as

$T_C = (\theta_D/1.45)exp\{-1.04(1+\lambda)/[\lambda-\mu^*(1+0.62\lambda)]\}$, (2)

where, $\mu^*$ is the Coulomb pseudo potential and $\theta_D$ is the Debye temperature. The λ is the electron-phonon coupling constant and is given by

$\lambda=N(E_F)<I^2>/M<\omega^2>$, (3)

where $N(E_F)$ is the DOS at the Fermi level, $<I^2>$ is the square averaged electronic matrix element for electron-phonon interaction, M is the ionic mass and $<\omega^2>$ is the square averaged phonon frequency. It appears from Eq. (2) that the change of λ and $\theta_D$ by pressure will determine the sign of $dT_C/dP$. It is well established that the pressure induces the lattice stiffening and generally reduces the $T_C$.[14-17] However, the DOS effect can either enhance or reduce the $T_C$ correspondingly by the increase or decrease of $N(E_F)$ due to applied pressure.[14-15] The dependence of $T_C$ on $\theta_D$ is complicated as it appears both in the linear and exponent (being connected with $<\omega^2>$ in Eq. (3)) terms of Eq.(2). Again, the change of exponent λ in Eq. (2) will be more effective than that of the linear term $\theta_D$ in determining $T_C$. The $\theta_D$ is generally increased by P amplifying the phonon frequency[19] as $<\omega^2> = 0.5\theta_D^2$ and thus may decrease λ (Eq. (3)) which in turn may reduce the $T_C$ (Eq.(2)). Therefore, the positive $dT_C/dP$ for $MgC_xNi_3$ is possibly originated from the increase of $N(E_F)$ and consequently by the enhancement of electron-phonon coupling constant λ (Eqs. (3)) if $\mu^*$ and $<I^2>$ are less pressure dependent. In addition, the P causes not only a shifting of the $E_F$ but also a broadening of the energy bands. This broadening of energy bands may also increase the $N(E_F)$. Most recently, Louis and Iyakutti[19] have successfully calculated the pressure effects on the $T_C$ of vanadium (V). Similarly, the computation of some important parameters such as $d\ln N(E_F)/dP$ and $d\ln\omega/dP$ of $MgC_xNi_3$ may be useful for quantitative analysis of our data.

Even though the strong spin fluctuations are unfavorable to exist in $MgC_xNi_3$,[21] the marginal or unstable spin fluctuations suppressing $T_C$ have not been totally ruled out.[8] In general, the pressure reduces the spin fluctuations and increases the $T_C$ because the spin fluctuations and superconductivity are mutually competitive phenomena. This may also be one of the reasons for the positive pressure effect on the $T_C$ of $MgC_xNi_3$. Another considerable factor showing a positive $dT_C/dP$ is the carbon stoichiometry in the sample. Generally, the deficiency of carbon from the optimum value decreases the $T_C$.[2,24] The non-stoichiometry of carbon (if any) may also affect the energy bands of the sample and alter the position of $E_F$ compared to that expected from theoretical energy band calculations[7-9] for stoichiometric $MgCNi_3$. The present investigations for three samples with different carbon content and $T_C$ show almost the same positive value of $dT_C/dP$ suggesting that the carbon deficiency does not significantly affect the pressure effect on $T_C$ of $MgC_xNi_3$. However, it is noted that Kumary et al.[25] recently found a decrease of $T_C$ up to a pressure of 17 kbar and an increase of $T_C$ beyond this pressure using resistivity measurements. It may be possible to briefly explain these controversial results as followings. (1) The $T_C$ determined from the resistivity (transport property) is always higher than that from susceptibility and specific heat (bulk property) measurements.[2,20-21,25] This may suggest that a small amount of higher $T_C$ phase existing in the grain boundaries[4] superconducts through percolation effects. (2) The negative $dT_C/dP$ observed in Ref. 25 using resistivity measurements at low pressures may be due to the reduction of grain boundary effects by pressure. Once the pressure is applied high enough (~17 kbar) to overcome the grain boundary effect, the bulk superconductivity dominates and the positive $dT_C/dP$ is found as the same with our results using susceptibility measurements.

In summary, the pressure increases the $T_C$ of three intermetallic, non-oxide, and perovskite electron-type superconductors $MgC_xNi_3$. The magnitude of change rate $d\ln T_C/dP$ in $MgC_xNi_3$ is about the same order as that in $MgB_2$ and $RNi_2B_2C$ (R = rare earths), which lies in the range of that of conventional superconductors. The positive value of $dT_C/dP$ for three $MgC_xNi_3$ samples are almost the same and independent of various $T_C$ resulted from different carbon stoichiometry. The present results of positive $dT_C/dP$ of $MgC_xNi_3$ can be explained mainly by the increase of density of states by pressure.

This work was supported by National Science Council of Republic of China under contract Nos. NSC91-2112-M110-005 and NSC90-2112-M009-025.

**Table 1.** The superconducting transition temperature $T_C$ (determined from the midpoint of resistive transition for $MgC_xNi_3$) at ambient pressure, $dT_C/dP$ and $dlnT_C/dP$ for some metallic as well as intermetallic superconductors.

| Sample composition | $T_C$ (K) | $dT_C/dP$ ($10^{-2}$ K/kbar) | $dlnT_C/dP$ ($10^{-3}$/kbar) | Reference |
|---|---|---|---|---|
| $MgB_2$ | 38.6 | –8.0 | –2.07 | 12 |
| $MgB_2$ | 37.5 | –16.0 | –4.26 | 13 |
| $LuNi_2B_2C$ | 15.9 | +1.88 | +1.18 | 14 |
| Ta | 4.3 | –0.26 | –0.60 | 16 |
| V | 5.3 | +1.0 | +1.88 | 18,19 |
| $MgC_xNi_3$(A) | 6.9 | +1.55 | +2.24 | This work |
| $MgC_xNi_3$(B) | 7.4 | +1.34 | +1.81 | This work |
| $MgC_xNi_3$(C) | 7.9 | +1.52 | +1.92 | This work |

**FIGURE CAPTIONS**

Fig.1: Temperature (T) variation of resistivity (ρ) and thermoelectric power (S) for sample A at ambient pressure. The inset shows the resistivity (ρ) of the three samples A, B and C near $T_C$.

Fig.2: Variation of ac magnetic susceptibility ($\chi_{ac}$) of samples A, B and C near $T_C$ at various pressures (P).

Fig.3: Pressure (P) dependence of superconducting transition temperature ($T_C$) of samples A, B and C.

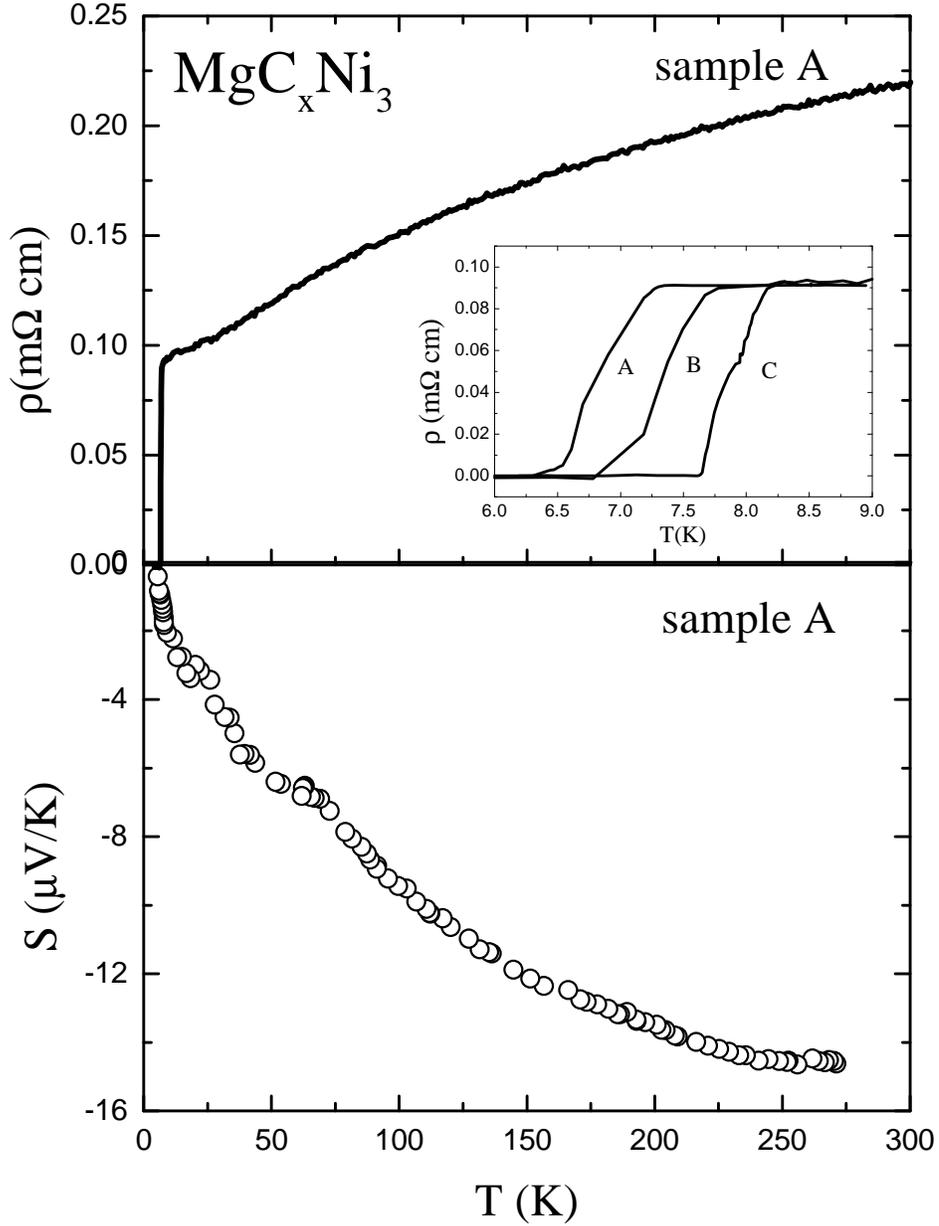

**Fig.1**  H. D. Yang *et al.*, PRB

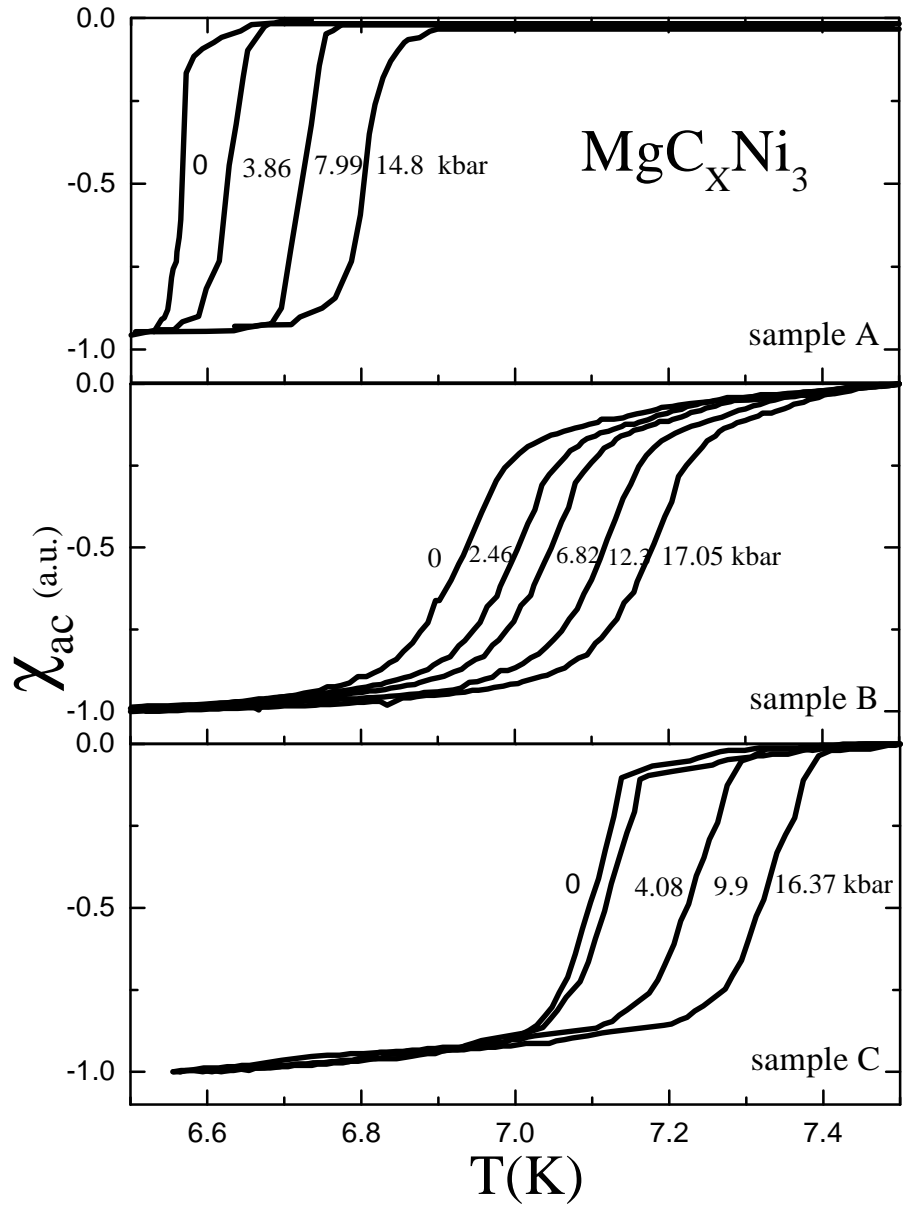

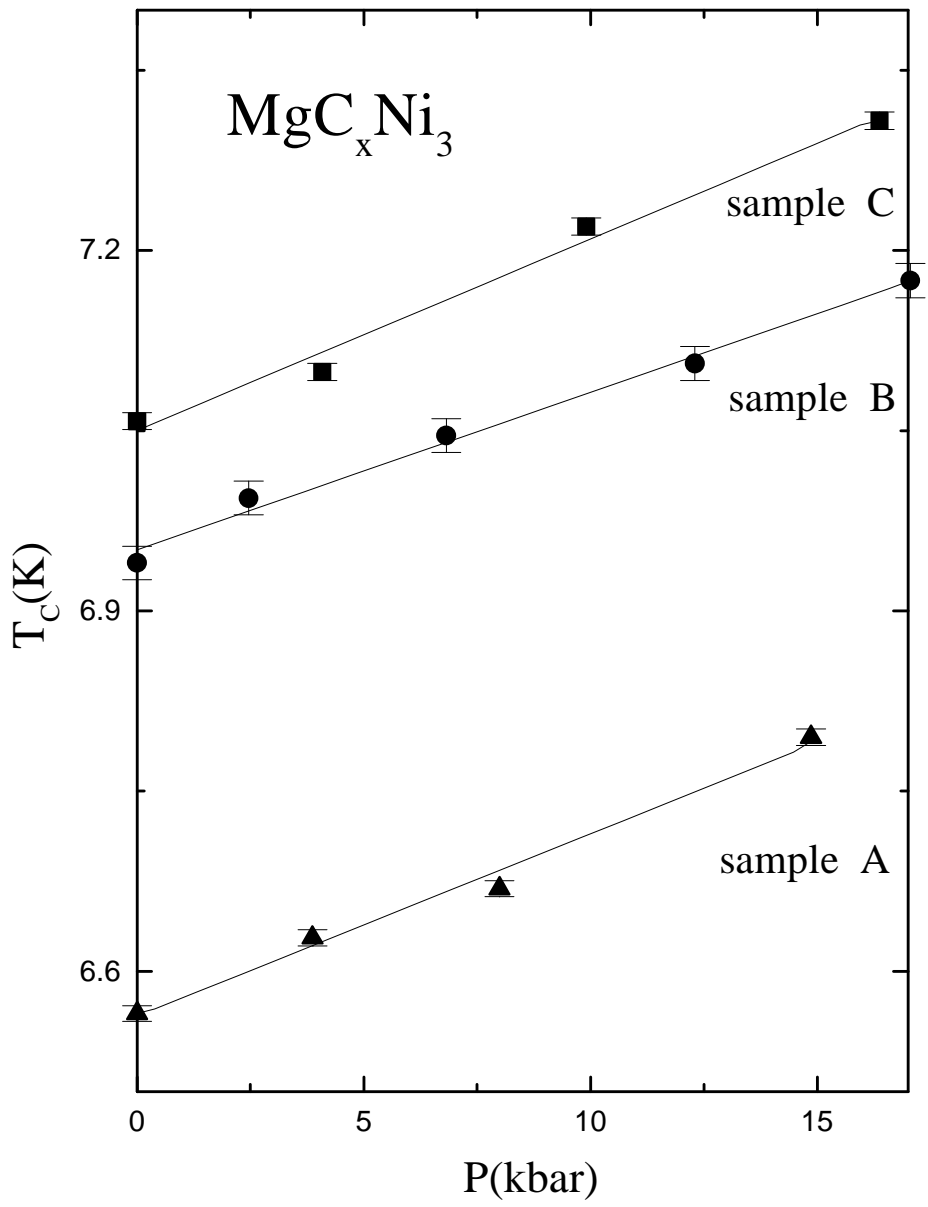

Fig.3            H. D. Yang *et al.*, PRB